\documentclass{achemso}
\usepackage[english]{babel}
\usepackage[utf8]{inputenc}
\usepackage[T1]{fontenc}
\usepackage{color}
\usepackage{rotating}
\usepackage{amssymb}
\usepackage{enumerate}
\usepackage{array,dcolumn}
\usepackage{xspace}
\usepackage{graphicx}
\usepackage{lscape}
\usepackage{rotating}
\usepackage{graphics}
\usepackage{amsmath}
\usepackage{amsfonts}
\usepackage{eso-pic}
\usepackage{epstopdf}
\usepackage{multimedia}
\usepackage{color}
\usepackage{times} 
\usepackage{bm}% bold math
\usepackage{txfonts} %solves non-visible Bold Greek latter in dvips
\usepackage{subcaption}
\newcommand{\C}{\text{c}}

\newcommand{\HF}{\text{HF}}

\newcommand{\SD}{\text{SD}}

\newcommand{\WC}{\text{c}_{II}}
\newcommand{\SC}{\text{c}_{I}}

\newcommand{\rr}{\mathbf{1}}
\newcommand{\rt}{\mathbf{2}}

\newcommand{\rn}{\mathbf{n}}

\newcommand{\drr}{d_{\mathbf{1}}}
\newcommand{\drt}{d_{\mathbf{2}}}
\newcommand{\drh}{d_{\mathbf{3}}}
\newcommand{\drn}{d_{\mathbf{n}}}

\newcommand{\ie}{{\it i.e.,\ }}

\author{Mireia Via-Nadal}
\affiliation{Donostia International Physics Center (DIPC),
20080 Donostia, Euskadi, Spain}
\alsoaffiliation
{Kimika Fakultatea, Euskal Herriko Unibertsitatea (UPV/EHU), Donostia, Euskadi, Spain}
\altaffiliation{ Equally contributed}
\author{Mauricio Rodr\'iguez-Mayorga}
\affiliation{Donostia International Physics Center (DIPC),
20080 Donostia, Euskadi, Spain}
\alsoaffiliation
{Kimika Fakultatea, Euskal Herriko Unibertsitatea (UPV/EHU), Donostia, Euskadi, Spain}
\alsoaffiliation
{Institut de Qu\'imica Computacional i Cat{\`a}lisi (IQCC) and
Departament de Qu\'imica, University of Girona, C/ Maria Aur\`elia Capmany, 69, 17003 Girona, Catalonia, Spain}
\altaffiliation{ Equally contributed}
\author{\\ Eloy Ramos-Cordoba}
\affiliation{Donostia International Physics Center (DIPC),
20080 Donostia, Euskadi, Spain}
\alsoaffiliation
{Kimika Fakultatea, Euskal Herriko Unibertsitatea (UPV/EHU), Donostia, Euskadi, Spain}
\author{Eduard Matito}
\affiliation{Donostia International Physics Center (DIPC),
20080 Donostia, Euskadi, Spain}
\alsoaffiliation
{IKERBASQUE, Basque Foundation for Science, 48013 Bilbao, Euskadi, Spain.}
\email{ematito@gmail.com}

\title{Singling Out Dynamic and Nondynamic Correlation}


\date{\today}

\begin{document}
\begin{abstract}
   The correlation part of the pair density is separated into two components, 
   one of them being predominant
   at short electronic ranges and the other at long ranges. The analysis of the intracular part
   of these components
   permits to classify molecular systems according to the prevailing correlation: dynamic or 
   nondynamic. The study of the long-range asymptotics reveals the key component of the pair
   density that is responsible for the description of London dispersion forces and a universal
   decay with the interelectronic distance.
   The natural range-separation, the identification of the dispersion forces and the kind 
   of predominant correlation type that arise from this analysis
   are expected to be important assets in 
   the development of new electronic structure methods in
   wavefunction, density and reduced density-matrix functional theories.
\end{abstract}

\newpage

\section{Introduction}
Electron correlation being the holy grail of electronic structure methods,
it has been the subject of extended 
analysis.~\cite{lowdin:55pr,cioslowski:91pra,gottlieb:05prl,raeber:15pra,benavides:17pra,
valderrama:97jcp,valderrama:99jcp,mok:96jpc,benavides:17pccp,ziesche:00the,ramos-cordoba:16pccp,ramos-cordoba:17jctc,juhasz:06jcp,mazziotti:98cpl}
The solution of quantum many-body problems hinges on the type of 
correlation present in the system, and
one of the most practical classifications consists
in the separation between dynamic- and nondynamic-correlation-including methods.
Indeed, there are accurate methods to study systems with one predominant
correlation type, but 
systems presenting both correlation types pose
one of the greatest current challenges in electronic structure 
theory.\cite{ramos-cordoba:14jcp,cioslowski:15jcp,pastorczak:17jctc}

The attempt at taking the best of both worlds has led
to a resurgence of interest in \textit{hybrid schemes},\cite{savin:88ijqc}
merging methods that recover different correlation types.\cite{grimme:99jcp,bao:17pccp,piris:17prl}
Among hybrid implementations, the most successful one is based on the range separation
of electron correlation,\cite{savin:88ijqc,iikura:01jcp,toulouse:09prl} using a mixing
function to combine approximations that 
account for short-range dynamic correlation ---such as density functional approximations---
with approaches providing correct long-range asymptotics. The performance of these methods
pivots on the choice of the function combining the two approaches, which provides a natural splitting
of the Coulomb interaction and thus the pair density.\cite{toulouse:04pra} 
In range-separation approximations, the typical choice is the error function that, in turn, depends on
an attenuating parameter, which is both system- and property-dependent.\cite{baer:10arpc,garrett:14jctc}
Even though the methods are chosen according to their ability of recovering dynamic and nondynamic
correlation, the range-separation of the pair density has not been motivated 
by the correlation type present in the system, risking double counting of electron correlation.

Thus far, 
there has been very few attempts to separate dynamic and nondynamic 
correlation,\cite{cioslowski:91pra,valderrama:97jcp,valderrama:99jcp,mok:96jpc,benavides:17pccp,vuckovic:17pccp,raeber:15pra,benavides:17pra,juhasz:06jcp}
most of them based on energy calculations.
The lack of a physically sound separation of dynamic and nondynamic correlation
precludes individual treatment of these effects.
We analyze the decomposition of the pair density 
into three components: the uncorrelated reference and two correlation terms.
The latter two behave differently with respect to large changes of 
the first-order reduced density matrix (1-RDM),
permitting the identification 
of systems with prevalent dynamic or nondynamic 
correlation.\cite{cioslowski:91pra,valderrama:97jcp,valderrama:99jcp,ramos-cordoba:16pccp,ramos-cordoba:17jctc}
Some of us have recently used a similar strategy to obtain scalar~\cite{ramos-cordoba:16pccp}
and local~\cite{ramos-cordoba:17jctc} measures of dynamic and nondynamic 
electron correlation from a two-electron model.
The intracule of the correlation
components of the pair density yields a two-fold separation of the Coulomb hole in terms 
of correlation type and interelectronic range. 
These components of the pair density
display a simpler mathematical form than the total pair density, 
one of them being dominant at short ranges and one with prevailing 
long-range contributions. This feature is particularly convenient for the design of energy functionals
in wavefunction, density and density matrix functional theories.
As a result of this separation, we will clearly identify the part of the pair density
that is responsible for the correct description of van der Waals interactions and unveil a universal
condition it should satisfy.\cite{via-nadal:17pra} To our knowledge, the latter is the only known
condition of the pair density
that can be employed to design methods including van der Waals interactions.

\section{Theoretical background}
\begin{figure}[h]
\centering
\includegraphics[scale=0.080]{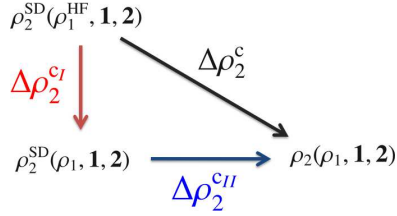}
\caption{The correlation part of the pair density, $\Delta \rho_2^{\C}$, decomposed into
   two components.}
\label{scheme}
\end{figure}

Let us consider the pair density of a $N$-electron system described by 
the $\Psi(\rr,\ldots,\rn)$ wavefunction,
\begin{eqnarray}
\rho_2(\rr,\rt)=\frac{N(N-1)}{2}\int\drh\ldots\drn\left|\Psi(\rr,\ldots,\rn)\right|^2 \,,
\end{eqnarray}
where numerical variables ($\rr,\rt,\ldots$) refer to space and spin coordinates.
Upon integration over its coordinates, the pair density can be reduced to the 
\textit{intracule density}, which only depends on the interelectronic range separation, $s$,
\begin{eqnarray}\label{eq:intrac}
I({\rho_2},s)= \int\drr\drt\, \rho_2(\rr,\rt) \delta(s-r_{12})  \,,
\end{eqnarray}
where $r_{12}$ is the Euclidean distance between the electrons at $\rr$ and $\rt$.
The intracule density 
is the simplest function in terms of which we can
express the Coulomb interaction energy,

\begin{eqnarray}\label{eq:vee}
V_{ee}\left[I\right]=\int ds\, \frac{I({\rho_2},s)}{s} \,.
\end{eqnarray}
The electron correlation contents of the pair density can be determined by the difference
between the actual pair density and an uncorrelated reference, which here we choose to be
the Hartree-Fock (HF) one,
\begin{equation}\label{eq:c}
\Delta \rho^{\C}_2(\rr,\rt)=\rho_2(\rr,\rt)-\rho_2^{\HF}(\rr,\rt) \,.
\end{equation}
The intracule of this function is Coulson's Coulomb hole,\cite{coulson:61ppsl}
\begin{equation}\label{eq:hc}
h_{\C}(s)=I({\Delta \rho_2^{\C}},s)=
\int\drr\drt\, \Delta \rho_2^{\C}(\rr,\rt) \delta(s-r_{12}) \,.
\end{equation}
In order to split the correlation part of the pair density (Eq.~\ref{eq:c}) we employ an approximate
pair density, the single-determinant (SD) ansatz of 
the pair density,\cite{lowdin:55pr} 
\begin{eqnarray}\label{eq:sd}
\rho_2^{\SD}(\rho_1,\rr,\rt)=\rho_1(\rr)\rho_1(\rt)-\left|\rho_1(\rr;\rt)\right|^2 \,,
\end{eqnarray}
where 
$\rho_1(\rr;\rt)$ is the 1-RDM and $\rho_1(\rr)\equiv\rho_1(\rr;\rr)$ is the
electron density. 
Substituting $\rho_1$ by the HF 1-RDM in Eq.~\ref{eq:sd}, 
yields the HF pair density, \ie 
\begin{equation}\label{eq:hf}
\rho_2^{\HF}(\rr,\rt)=\rho_2^{\SD}(\rho_1^{\HF},\rr,\rt)  \,, 
\end{equation}
which does not account for electron correlation.
However, $\rho_2^{\SD}(\rho_1,\rr,\rt)$ 
can be regarded as an approximation to the
actual pair density; an approximation which
does not account for dynamic correlation
either at short~\cite{rodriguez:17pccp2}
or at long range.\cite{via-nadal:17pra} Figure \ref{scheme} depicts the two paths of
arriving at the exact $\rho_2(\rho_1,\rr,\rt)$ from $\rho_2^{\SD}(\rho_1^{\HF},\rr,\rt)$, 
either straightforwardly or 
through the intermediate SD approximation. The latter path defines the decomposition
of the correlation part of the pair density, 
\begin{eqnarray}\label{eq:part}\nonumber
   \Delta \rho_2^{\C}(\rr,\rt) &=&
   \left(\rho_2(\rho_1,\rr,\rt)-\rho_2^{\SD}(\rho_1,\rr,\rt)\right)
   +\left(\rho_2^{\SD}(\rho_1,\rr,\rt)-\rho_2^{\SD}(\rho_1^{\HF},\rr,\rt)\right)  \\
   &=&
   \Delta \rho_2^{\SC}(\rr,\rt)+\Delta \rho_2^{\WC}(\rr,\rt) \,. 
\end{eqnarray}

$\Delta\rho_2^{\SC}(\rr,\rt)$ will be large only if the HF 1-RDM and the actual 1-RDM
are significantly different and, in such case, the system will be affected by
nondynamic correlation. Indeed, the wavefunction of systems dominated by dynamic 
correlation can be described by a large expansion of Slater determinants with one 
of them (the HF one) having an expansion coefficient very close to one.\cite{Note42} 
Therefore, these systems are
characterized by a 1-RDM that retains the shape of the HF 1-RDM. Conversely, 
the wavefunction of nondynamic-correlated systems can be written as a shorter 
expansion of Slater determinants, but in this case the HF determinant has
an expansion coefficient that is qualitatively smaller than one.\cite{Note42}
Since the 1-RDM
is determined by the square of the expansion coefficients, we expect systems affected
by nondynamic correlation to display large $\Delta\rho_2^{\SC}(\rr,\rt)$.
Some authors have used similar arguments to use the electron density
(the diagonal part of the 1-RDM) as a means to define dynamic and nondynamic correlation 
energy.\cite{cioslowski:91pra,valderrama:97jcp} 
In this work, we prefer to employ the 1-RDM because the cases of spin entaglement
would not be regarded as nondynamic correlation if only density differences were considered.
Indeed, in the stretched H$_2$ molecule, the HF electron density is qualitatively
similar to the exact one, whereas there are large and notorious differences between
the exact and the HF 1-RDMs.\newline

The magnitude of $\Delta\rho_2^{\SC}(\rr,\rt)$ can be thus
regarded as a measure of nondynamic correlation but it can also be interpreted as
the correlation retrieved by using the actual 1-RDM rather than the HF one 
to construct the pair density.
Conversely, $\Delta\rho_2^{\WC}(\rr,\rt)$ does not depend on the differences 
between $\rho_1^{\HF}$ and $\rho_1$, but on the validity of the SD approximation.
Note that 
$\Delta \rho_2^{\WC}$ coincides with the cumulant of the pair density.\cite{kutzelnigg:99jcp,mazziotti:98cpl}
The intracule functions of $\rho_2^{\SD}(\rho_1,\rr,\rt)$ and the exact
pair density, $\rho_2(\rho_1,\rr,\rt)$, display the same asymptotic behavior~\cite{ernzerhof:96jcp} and,
therefore, $\Delta \rho_2^{\WC}$ is dominated by the short-range component. 
Interestingly, $\Delta\rho_2^{\SC}(\rr,\rt)$ is the long-range-dominant component
of the correlated part of the pair density (Eq.~\ref{eq:c}) because the HF and
the exact 1-RDM can differ substantially at large separations, for instance, in
the presence of entanglement. On the contrary, $\Delta\rho_2^{\SC}(\rr,\rt)$ displays
very small values at small interelectronic distances mostly due to the opposite-spin
part of this term.\newline

The current partition, 
\begin{equation}\label{eq:part}
\rho_2(\rr,\rt)=\rho_2^{\HF}(\rr,\rt)+
\Delta \rho_2^{\SC}(\rr,\rt)+\Delta \rho_2^{\WC}(\rr,\rt) \,,
\end{equation}
provides a natural range separation of the pair density that
can be employed to split the Coulomb hole into two correlation
components,
\begin{equation}
h_{\C}(s)=h_{\SC}(s)+h_{\WC}(s)=I(\Delta \rho_2^{\SC},s)+I(\Delta \rho_2^{\WC},s) \,,
\end{equation}
naturally yielding a separation of electron correlation by range.
We will show that the decay of $I(\Delta \rho_2^{\WC},R)$ is universal
and it corresponds to a characteristic signature of London dispersion forces
($R$ being the distance between two atoms in the molecule).

\section{Results and Discussion}
In the following we introduce five selected 
examples that illustrate the effectiveness of the current scheme to separate
the correlation part of the Coulomb hole
at different ranges and how the long-range of $\Delta\rho_2^{\WC}$
can be used to identify and characterize van der Waals interactions.

\begin{figure}[h]
\centering
\includegraphics[scale=0.68]{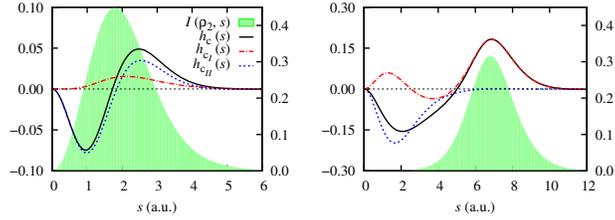}
\caption{The total Coulomb hole (black), $h_{\SC}(s)$ (red) and $h_{\WC}(s)$ 
(blue) correlation components, and the intracule density (shadowed green region, right $y$-axis) of 
the H$_2$ molecule at $1.32$ and $7.56$ a.u. bond lengths.}
\label{h2}
\end{figure}

\textit{The Hydrogen Molecule}\cite{Note41}.---
At the equilibrium geometry, $h_{\WC}(s)$ dominates over
$h_{\SC}(s)$ at all interelectronic distances $s$, as shown in the left panel of Fig.~\ref{h2},
whereas $h_{\SC}(s)$ increases importantly as the bond is stretched, in line with
the expected increase of nondynamic correlation.
The most likely distribution of the electron pair at large bond lengths corresponds to one
electron sitting at each atom and, accordingly, the intracule density peaks around the
bond-length distance. 
At the dissociation limit, the long-range part of 
the Coulomb hole is completely determined by $h_{\SC}(s)$
because one isolated electron cannot give rise to
dynamic correlation. Hence, the unrestricted HF calculation
of H$_2$ produces Coulomb hole components that are not distinguishable from 
FCI.\cite{mercero:03bch}
A simple interpretation is also obtained from valence bond theory: at large separations,
the exact pair density is entirely described by covalent components, whereas the HF pair
density contains equally contributing ionic and covalent terms. $h_{\WC}(s)$
removes the ionic contribution (\textit{i.e.}, removes contributions keeping the electrons
at short distances), whereas $h_{\SC}(s)$ adds the missing covalent
contribution (\textit{i.e.}, adds contributions placing one electron in each
atom); in accord with the results plotted in the r.h.s. of Fig.~\ref{h2} 
(see also Supp. Material).

\begin{figure}[h]
\centering
\includegraphics[scale=0.68]{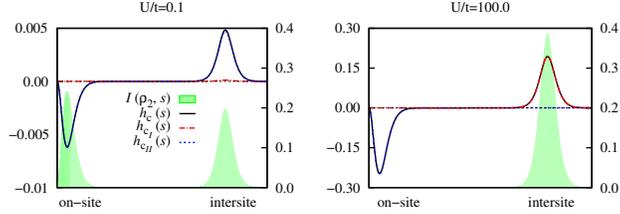}
\caption{
The Coulomb hole (see Fig.~\ref{h2} for further details) of 
the two-site real-space Hubbard model for various $U/t$ values.}
\label{hubb}
\end{figure}

\textit{The Hubbard Dimer}.--- 
The Hubbard dimer is the simplest model of interacting particles in a lattice
and conceivably the most studied model for testing methods at different
regimes.\cite{carrascal:15jpcm,deur:17prb} We employ
the one-dimension Hamiltonian of the Hubbard model,
\begin{equation}
\hat{H}=
-t\sum_{\left<\mu,\nu\right>,\sigma}\left(\hat{c}^{\dagger}_{\mu\sigma}\hat{c}_{\nu\sigma}
   +\hat{c}^{\dagger}_{\nu\sigma}\hat{c}_{\mu\sigma}\right)
   +U\sum_{\mu}\hat{\rho}_{\mu\alpha}\hat{\rho}_{\mu\beta} \,,
\end{equation}
where $\mu$ and $\nu$ denote the sites, $\sigma$ the spin polarization ($\alpha$
or $\beta$), $\hat{c}^{\dagger}_{\mu\sigma}$ and $\hat{c}_{\mu\sigma}$ are creation and 
annihilation operators of one electron with spin $\sigma$ in site $\mu$, and
$\hat{\rho}_{\mu\sigma}$ stands for a one-particle number operator with spin $\sigma$
acting on site $\mu$. $t$ is the hopping parameter and $U$ 
is the on-site interaction parameter. 
These parameters control the electron correlation within the Hubbard model, 
small (large) $U/t$ inducing dynamic (nondynamic) correlation. Hence large $U/t$
values prompt the electrons to distribute among the sites 
to minimize the electron repulsion.
Fig.~\ref{hubb}
presents plots of the Coulomb hole at 
various values of $U/t$ for the two-electron two-site Hubbard model in real 
space.\cite{carrascal:15jpcm}
At low $U/t$ values, the system is barely affected by correlation,
thus dynamic correlation dominates (small $h_{\SC}(s)$ and large $h_{\WC}(s)$)
and the electron pairs distribute
equally between on-site and intersite components.
As $U/t$ grows, nondynamic correlation dominates and $h_{\SC}(s)$
becomes more important, being the prevailing contribution between sites.

\begin{figure}[h]
\centering
\includegraphics[scale=0.485]{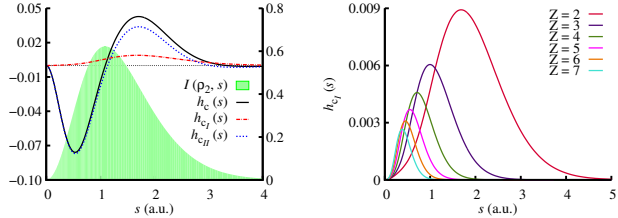}
\caption{
The Coulomb hole of He (l.h.s.) and
$h_{\SC}(s)$ 
for the isoelectronic series of He (r.h.s.).}
\label{hez}
\end{figure}

\textit{The He series}.\cite{Note41}--- 
The He isoelectronic series is perhaps the simplest series of systems dominated
by dynamic correction.\cite{chakravorty:93pra} 
As the atomic number $Z$ increases, the electron correlation of He($Z$)
tends to a constant and the exact electron density barely distinguishes 
from the HF one. In Fig.~\ref{hez} we observe that 
$h_{\SC}(s)$
decreases with the atomic number $Z$ and, hence,
$h_{\WC}$ completely takes over.

\textit{$N$ hydrogen atoms}.\cite{Note41}--- 
The size consistency of our approach and its
ability to measure spin entanglement is examined in Fig.~\ref{hn}. 
We have plotted the Coulomb hole of the $N$-vertex polyhedron resulting
from $N$ hydrogen atoms separated by 10$\AA\,$ from the center of the
polyhedron. At these large separations, the hydrogen atoms only interact 
to each other through entanglement and this is the only term that remains
in the cumulant,\cite{raeber:15pra} (\textit{i.e.} in $\Delta \rho_2^{\WC}$), which shows a linear 
behavior with $N$ (see Fig.~\ref{hn}). As in previous systems, 
$h_{\WC}$ is short ranged and its contribution to the energy grows
linearly with $N$. These systems can be classified as nondynamic correlated
because $h_{\SC}$ is mostly long ranged and peaks at the same positions
of the intracule density maxima.
The planar $D_{4h}/D_{2h}$ potential energy surface of H$_4$ has also been 
used for discriminating between dynamic and nondynamic correlation~\cite{ramos-cordoba:15jcp} 
and is given in the Supp. Material.

{
\begin{figure}[h]
\centering
\includegraphics[scale=0.71]{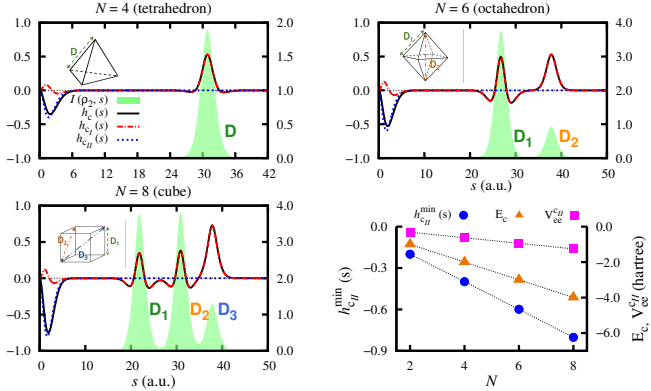}
\caption{
The Coulomb hole of
$N$ hydrogen atoms.  D, D$_1$, D$_2$, and D$_3$ 
indicate the different distances between the H atoms placed at the vertices of the 
respective polyhedra.
The bottom r.h.s. plot displays
the minimal value of $h_{\WC}$, $h_{\WC}^{\text{min}}(s)$,
the part of electron-electron repulsion that corresponds to $\Delta \rho_2^{\WC}$
(\textit{i.e.} $V_{ee}^{\WC}=\int ds\, h_{\WC}(s)/s$), and the correlation energy (E$_{c}$),
as a function of $N$.}
\label{hn}
\end{figure}
}

\begin{figure}[h]
\centering
\includegraphics[scale=0.71]{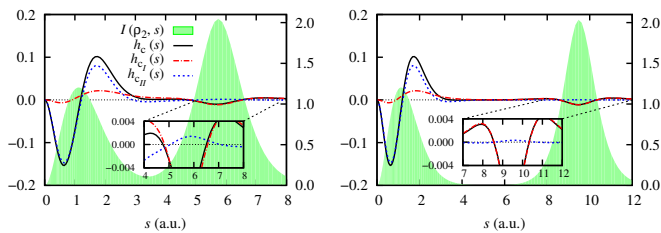}
\caption{
The Coulomb hole of
the He$_2$ molecule at two bond lengths ($5.67$ and $9.45$ a.u., left and right).
The inset plots reproduce the ones above on a narrower interval.}
\label{he2}
\end{figure}

\begin{figure}[h]
\centering
\includegraphics[scale=0.54]{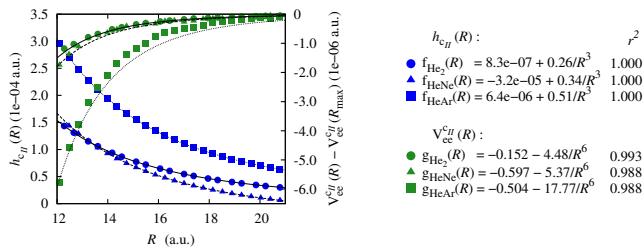}
\caption{$h_{\WC}(R)$ (in blue) against the bond length ($R$)
and the corresponding electron-electron van der Waals contributions (in green)
for several noble-gas dimers: He$_2$, HeNe and HeAr.}
\label{he2corr}
\end{figure}

\textit{van der Waals (vdW) Interactions}.\cite{Note41}---
Fig.~\ref{he2} includes plots of the Coulomb hole of the helium dimer.
$h_c$ compares satisfactorily to earlier calculations.\cite{piris:08jcp2}
The dynamic long-range interaction between the two noble-gas atoms is reflected by the
second peak of the intracule density, whereas the interaction of the electron pair
within each helium shows in the first peak.
Regardless the bond length, $h_{\WC}$ dominates, indicating
that the correlation is dynamic and mainly affects the electron pair within each He.
Unlike H$_2$, there is very little long-range nondynamic correlation in this system;
however, at all distances, the long-range part of $h_{\WC}$
peaks around the bond-length distance (see the inset plots of Fig.~\ref{he2}).
The plot in Fig.~\ref{he2corr} presents $h_{\WC}(R)$ against the bond length, $R$,
revealing a $R^{-3}$ decay.
It is a textbook fact that the pairwise vdW energy decays like $R^{-6}$.\cite{pauling:35book}
Using perturbation theory, we have recently proved that 
the vdW contribution to $h_{\WC}$ should actually decay like $R^{-3}$,
the integration of $h_{\WC}(s)/s$ over $s$ yielding a fraction of the Coulombic
interaction (Eq.~\ref{eq:vee}) due to London dispersion forces and, therefore, 
decaying as $R^{-6}$.\cite{via-nadal:17pra}
Fig.~\ref{he2corr} includes plots for other noble-gas dimers, which also satisfy this
property. 
Most density functional theory (DFT) practitioners add 
\textit{ad hoc} empirical corrections to the energy 
for vdW interactions and, therefore, they only shift the relative energies of
different conformers, yet the electronic structure of the system is not 
completely considered.\cite{hermann:17cr}
The present separation into correlation regimes unveils the target part of the pair 
density and the Coulomb hole, 
\textit{i.e.}, the long-range component of $h_{\WC}(s)$,
which should be improved in order to 
incorporate the description of London dispersion forces and avoid the latter problem,
thus opening a door to the accurate account of these forces within DFT and 
reduced density matrix functional theory (RDMFT).

In conclusion, Eqs. 8 and 9 represent a separation 
of the pair density and the Coulomb hole into 
components dominated by short- and long-range interactions.
This result is expected to be important in the development of new hybrid electronic 
structure methods that can be employed in RDMFT~\cite{piris:14ijqc,piris:17prl}
and other computational approaches. 
For instance, the HF reference in Fig.~1 can be
replaced by the Kohn-Sham system to adapt the present idea to DFT.
It can be shown that 
the exchange-correlation functional can be entirely written in terms of the
Kohn-Sham orbitals, $\Delta\rho_2^{\SC}$, and $\Delta\rho_2^{\WC}$.
Hence, a template to construct density functional approximations, where 
the correlation components are treated separately, arises.
Such possibility is already being explored in our laboratory.

%%-----------------------------------------------------

\section*{Acknowledgements}
We thank D. Casanova, X. Lopez, P. Salvador, and specially, P.M.W. Gill and J.M. Ugalde for helpful insights.
This research has been funded by Spanish MINECO/FEDER Projects 
CTQ2014-52525-P, PGC2018-098212-B-C21, and EUIN2017-88605.
We acknowledge doctoral grants BES-2015-072734 and FPU-2013/00176,
and the funding from the European Union's
Horizon 2020 research and innovation programme under the
Marie Sklodowska-Curie grant agreement (No. 660943).\newline

%${^{\dagger}}$M.V.N. and M.R.M. contributed equally to this work.
\textbf{Supporting Information Available}: Analysis of the $D_{2h}/D_{4h}$
H$_4$ planar molecule and the hydrogen molecule
   from ionic and covalent contributions. Full plots of H$_2$, the two-site Hubbard dimer, the He isoelectronic series, and the $N$ hydrogen atoms model. 

%------------------------------------------------
% References
%------------------------------------------------
\bibliography{gen,jpcl_notes}

\providecommand{\latin}[1]{#1}
\makeatletter
\providecommand{\doi}
  {\begingroup\let\do\@makeother\dospecials
  \catcode`\{=1 \catcode`\}=2 \doi@aux}
\providecommand{\doi@aux}[1]{\endgroup\texttt{#1}}
\makeatother
\providecommand*\mcitethebibliography{\thebibliography}
\csname @ifundefined\endcsname{endmcitethebibliography}
  {\let\endmcitethebibliography\endthebibliography}{}
\begin{mcitethebibliography}{45}
\providecommand*\natexlab[1]{#1}
\providecommand*\mciteSetBstSublistMode[1]{}
\providecommand*\mciteSetBstMaxWidthForm[2]{}
\providecommand*\mciteBstWouldAddEndPuncttrue
  {\def\EndOfBibitem{\unskip.}}
\providecommand*\mciteBstWouldAddEndPunctfalse
  {\let\EndOfBibitem\relax}
\providecommand*\mciteSetBstMidEndSepPunct[3]{}
\providecommand*\mciteSetBstSublistLabelBeginEnd[3]{}
\providecommand*\EndOfBibitem{}
\mciteSetBstSublistMode{f}
\mciteSetBstMaxWidthForm{subitem}{(\alph{mcitesubitemcount})}
\mciteSetBstSublistLabelBeginEnd
  {\mcitemaxwidthsubitemform\space}
  {\relax}
  {\relax}

\bibitem[L{\"o}wdin(1955)]{lowdin:55pr}
L{\"o}wdin,~P.-O. Quantum theory of many-particle systems. I. Physical
  interpretations by means of density matrices, natural spin-orbitals, and
  convergence problems in the method of configurational interaction.
  \emph{Phys. Rev.} \textbf{1955}, \emph{97}, 1474--1489\relax
\mciteBstWouldAddEndPuncttrue
\mciteSetBstMidEndSepPunct{\mcitedefaultmidpunct}
{\mcitedefaultendpunct}{\mcitedefaultseppunct}\relax
\EndOfBibitem
\bibitem[Cioslowski(1991)]{cioslowski:91pra}
Cioslowski,~J. Density-driven self-consistent-field method: Density-constrained
  correlation energies in the helium series. \emph{Phys. Rev. A} \textbf{1991},
  \emph{43}, 1223--1228\relax
\mciteBstWouldAddEndPuncttrue
\mciteSetBstMidEndSepPunct{\mcitedefaultmidpunct}
{\mcitedefaultendpunct}{\mcitedefaultseppunct}\relax
\EndOfBibitem
\bibitem[Gottlieb and Mauser(2005)Gottlieb, and Mauser]{gottlieb:05prl}
Gottlieb,~A.~D.; Mauser,~N.~J. New measure of electron correlation. \emph{Phys.
  Rev. Lett.} \textbf{2005}, \emph{95}, 123003\relax
\mciteBstWouldAddEndPuncttrue
\mciteSetBstMidEndSepPunct{\mcitedefaultmidpunct}
{\mcitedefaultendpunct}{\mcitedefaultseppunct}\relax
\EndOfBibitem
\bibitem[Raeber and Mazziotti(2015)Raeber, and Mazziotti]{raeber:15pra}
Raeber,~A.; Mazziotti,~D.~A. Large eigenvalue of the cumulant part of the
  two-electron reduced density matrix as a measure of off-diagonal long-range
  order. \emph{Phys. Rev. A} \textbf{2015}, \emph{92}, 052502\relax
\mciteBstWouldAddEndPuncttrue
\mciteSetBstMidEndSepPunct{\mcitedefaultmidpunct}
{\mcitedefaultendpunct}{\mcitedefaultseppunct}\relax
\EndOfBibitem
\bibitem[Benavides-Riveros \latin{et~al.}(2017)Benavides-Riveros, Lathiotakis,
  Schilling, and Marques]{benavides:17pra}
Benavides-Riveros,~C.~L.; Lathiotakis,~N.~N.; Schilling,~C.; Marques,~M.~A.
  Relating correlation measures: The importance of the energy gap. \emph{Phys.
  Rev. A} \textbf{2017}, \emph{95}, 032507\relax
\mciteBstWouldAddEndPuncttrue
\mciteSetBstMidEndSepPunct{\mcitedefaultmidpunct}
{\mcitedefaultendpunct}{\mcitedefaultseppunct}\relax
\EndOfBibitem
\bibitem[Valderrama \latin{et~al.}(1997)Valderrama, Lude{\~n}a, and
  Hinze]{valderrama:97jcp}
Valderrama,~E.; Lude{\~n}a,~E.~V.; Hinze,~J. {Analysis of dynamical and
  nondynamical components of electron correlation energy by means of
  local-scaling density-functional theory}. \emph{J. Chem. Phys.}
  \textbf{1997}, \emph{106}, 9227--9235\relax
\mciteBstWouldAddEndPuncttrue
\mciteSetBstMidEndSepPunct{\mcitedefaultmidpunct}
{\mcitedefaultendpunct}{\mcitedefaultseppunct}\relax
\EndOfBibitem
\bibitem[Valderrama \latin{et~al.}(1999)Valderrama, Lude{\~n}a, and
  Hinze]{valderrama:99jcp}
Valderrama,~E.; Lude{\~n}a,~E.~V.; Hinze,~J. {Assessment of dynamical and
  nondynamical correlation energy components for the beryllium-atom
  isoelectronic sequence}. \emph{J. Chem. Phys.} \textbf{1999}, \emph{110},
  2343--2353\relax
\mciteBstWouldAddEndPuncttrue
\mciteSetBstMidEndSepPunct{\mcitedefaultmidpunct}
{\mcitedefaultendpunct}{\mcitedefaultseppunct}\relax
\EndOfBibitem
\bibitem[Mok \latin{et~al.}(1996)Mok, Neumann, and Handy]{mok:96jpc}
Mok,~D. K.~W.; Neumann,~R.; Handy,~N.~C. Dynamic and Nondynamic Correlation.
  \emph{J. Phys. Chem.} \textbf{1996}, \emph{100}, 6225--6230\relax
\mciteBstWouldAddEndPuncttrue
\mciteSetBstMidEndSepPunct{\mcitedefaultmidpunct}
{\mcitedefaultendpunct}{\mcitedefaultseppunct}\relax
\EndOfBibitem
\bibitem[Benavides-Riveros \latin{et~al.}(2017)Benavides-Riveros, Lathiotakis,
  and Marques]{benavides:17pccp}
Benavides-Riveros,~C.~L.; Lathiotakis,~N.~N.; Marques,~M.~A. Towards a formal
  definition of static and dynamic electronic correlations. \emph{Phys. Chem.
  Chem. Phys.} \textbf{2017}, \emph{19}, 12655--12664\relax
\mciteBstWouldAddEndPuncttrue
\mciteSetBstMidEndSepPunct{\mcitedefaultmidpunct}
{\mcitedefaultendpunct}{\mcitedefaultseppunct}\relax
\EndOfBibitem
\bibitem[Ziesche(2000)]{ziesche:00the}
Ziesche,~P. On relations between correlation, fluctuation and localization.
  \emph{J. Mol. Struct. (Theochem)} \textbf{2000}, \emph{527}, 35--50\relax
\mciteBstWouldAddEndPuncttrue
\mciteSetBstMidEndSepPunct{\mcitedefaultmidpunct}
{\mcitedefaultendpunct}{\mcitedefaultseppunct}\relax
\EndOfBibitem
\bibitem[Ramos-Cordoba \latin{et~al.}(2016)Ramos-Cordoba, Salvador, and
  Matito]{ramos-cordoba:16pccp}
Ramos-Cordoba,~E.; Salvador,~P.; Matito,~E. Separation of dynamic and
  nondynamic correlation. \emph{Phys. Chem. Chem. Phys.} \textbf{2016},
  \emph{18}, 24015--24023\relax
\mciteBstWouldAddEndPuncttrue
\mciteSetBstMidEndSepPunct{\mcitedefaultmidpunct}
{\mcitedefaultendpunct}{\mcitedefaultseppunct}\relax
\EndOfBibitem
\bibitem[Ramos-Cordoba and Matito(2017)Ramos-Cordoba, and
  Matito]{ramos-cordoba:17jctc}
Ramos-Cordoba,~E.; Matito,~E. Local Descriptors of dynamic and nondynamic
  correlation. \emph{J. Chem. Theory Comput.} \textbf{2017}, \emph{13},
  2705--2711\relax
\mciteBstWouldAddEndPuncttrue
\mciteSetBstMidEndSepPunct{\mcitedefaultmidpunct}
{\mcitedefaultendpunct}{\mcitedefaultseppunct}\relax
\EndOfBibitem
\bibitem[Juh{\'a}sz and Mazziotti(2006)Juh{\'a}sz, and Mazziotti]{juhasz:06jcp}
Juh{\'a}sz,~T.; Mazziotti,~D.~A. The cumulant two-particle reduced density
  matrix as a measure of electron correlation and entanglement. \emph{J. Chem.
  Phys.} \textbf{2006}, \emph{125}, 174105\relax
\mciteBstWouldAddEndPuncttrue
\mciteSetBstMidEndSepPunct{\mcitedefaultmidpunct}
{\mcitedefaultendpunct}{\mcitedefaultseppunct}\relax
\EndOfBibitem
\bibitem[Mazziotti(1998)]{mazziotti:98cpl}
Mazziotti,~D.~A. Approximate solution for electron correlation through the use
  of Schwinger probes. \emph{Chem. Phys. Lett.} \textbf{1998}, \emph{289},
  419--427\relax
\mciteBstWouldAddEndPuncttrue
\mciteSetBstMidEndSepPunct{\mcitedefaultmidpunct}
{\mcitedefaultendpunct}{\mcitedefaultseppunct}\relax
\EndOfBibitem
\bibitem[Ramos-Cordoba \latin{et~al.}(2014)Ramos-Cordoba, Salvador, Piris, and
  Matito]{ramos-cordoba:14jcp}
Ramos-Cordoba,~E.; Salvador,~P.; Piris,~M.; Matito,~E. Two new constraints for
  the cumulant matrix. \emph{J. Chem. Phys.} \textbf{2014}, \emph{141},
  234101\relax
\mciteBstWouldAddEndPuncttrue
\mciteSetBstMidEndSepPunct{\mcitedefaultmidpunct}
{\mcitedefaultendpunct}{\mcitedefaultseppunct}\relax
\EndOfBibitem
\bibitem[Cioslowski \latin{et~al.}(2015)Cioslowski, Piris, and
  Matito]{cioslowski:15jcp}
Cioslowski,~J.; Piris,~M.; Matito,~E. Robust validation of approximate 1-matrix
  functionals with few-electron harmonium atoms. \emph{J. Chem. Phys.}
  \textbf{2015}, \emph{143}, 214101\relax
\mciteBstWouldAddEndPuncttrue
\mciteSetBstMidEndSepPunct{\mcitedefaultmidpunct}
{\mcitedefaultendpunct}{\mcitedefaultseppunct}\relax
\EndOfBibitem
\bibitem[Pastorczak \latin{et~al.}(2017)Pastorczak, Shen, Hapka, Piecuch, and
  Pernal]{pastorczak:17jctc}
Pastorczak,~E.; Shen,~J.; Hapka,~M.; Piecuch,~P.; Pernal,~K. Intricacies of van
  der Waals interactions in systems with elongated bonds revealed by
  electron-groups embedding and high-level coupled-cluster approaches. \emph{J.
  Chem. Theory Comput.} \textbf{2017}, \emph{13}, 5404--5419\relax
\mciteBstWouldAddEndPuncttrue
\mciteSetBstMidEndSepPunct{\mcitedefaultmidpunct}
{\mcitedefaultendpunct}{\mcitedefaultseppunct}\relax
\EndOfBibitem
\bibitem[Savin(1988)]{savin:88ijqc}
Savin,~A. A combined density functional and configuration interaction method.
  \emph{Int. J. Quantum Chem.} \textbf{1988}, \emph{34}, 59--69\relax
\mciteBstWouldAddEndPuncttrue
\mciteSetBstMidEndSepPunct{\mcitedefaultmidpunct}
{\mcitedefaultendpunct}{\mcitedefaultseppunct}\relax
\EndOfBibitem
\bibitem[Grimme and Waletzke(1999)Grimme, and Waletzke]{grimme:99jcp}
Grimme,~S.; Waletzke,~M. A combination of Kohn--Sham density functional theory
  and multi-reference configuration interaction methods. \emph{J. Chem. Phys.}
  \textbf{1999}, \emph{111}, 5645--5655\relax
\mciteBstWouldAddEndPuncttrue
\mciteSetBstMidEndSepPunct{\mcitedefaultmidpunct}
{\mcitedefaultendpunct}{\mcitedefaultseppunct}\relax
\EndOfBibitem
\bibitem[Bao \latin{et~al.}(2017)Bao, Gagliardi, and Truhlar]{bao:17pccp}
Bao,~J.~J.; Gagliardi,~L.; Truhlar,~D.~G. Multiconfiguration pair-density
  functional theory for doublet excitation energies and excited state
  geometries: the excited states of CN. \emph{Phys. Chem. Chem. Phys.}
  \textbf{2017}, \emph{19}, 30089--30096\relax
\mciteBstWouldAddEndPuncttrue
\mciteSetBstMidEndSepPunct{\mcitedefaultmidpunct}
{\mcitedefaultendpunct}{\mcitedefaultseppunct}\relax
\EndOfBibitem
\bibitem[Piris(2017)]{piris:17prl}
Piris,~M. Global Method For The Electron Correlation. \emph{Phys. Rev. Lett.}
  \textbf{2017}, \emph{119}, 063002\relax
\mciteBstWouldAddEndPuncttrue
\mciteSetBstMidEndSepPunct{\mcitedefaultmidpunct}
{\mcitedefaultendpunct}{\mcitedefaultseppunct}\relax
\EndOfBibitem
\bibitem[Iikura \latin{et~al.}(2001)Iikura, Tsuneda, Yanai, and
  Hirao]{iikura:01jcp}
Iikura,~H.; Tsuneda,~T.; Yanai,~T.; Hirao,~K. A long-range correction scheme
  for generalized-gradient-approximation exchange functionals. \emph{J. Chem.
  Phys.} \textbf{2001}, \emph{115}, 3540--3544\relax
\mciteBstWouldAddEndPuncttrue
\mciteSetBstMidEndSepPunct{\mcitedefaultmidpunct}
{\mcitedefaultendpunct}{\mcitedefaultseppunct}\relax
\EndOfBibitem
\bibitem[Toulouse \latin{et~al.}(2009)Toulouse, Gerber, Jansen, Savin, and
  Angy{\'a}n]{toulouse:09prl}
Toulouse,~J.; Gerber,~I.~C.; Jansen,~G.; Savin,~A.; Angy{\'a}n,~J.~G.
  Adiabatic-connection fluctuation-dissipation density-functional theory based
  on range separation. \emph{Phys. Rev. Lett.} \textbf{2009}, \emph{102},
  096404\relax
\mciteBstWouldAddEndPuncttrue
\mciteSetBstMidEndSepPunct{\mcitedefaultmidpunct}
{\mcitedefaultendpunct}{\mcitedefaultseppunct}\relax
\EndOfBibitem
\bibitem[Toulouse \latin{et~al.}(2004)Toulouse, Colonna, and
  Savin]{toulouse:04pra}
Toulouse,~J.; Colonna,~F.; Savin,~A. Long-range short-range separation of the
  electron-electron interaction in density-functional theory. \emph{Phys. Rev.
  A} \textbf{2004}, \emph{70}, 062505\relax
\mciteBstWouldAddEndPuncttrue
\mciteSetBstMidEndSepPunct{\mcitedefaultmidpunct}
{\mcitedefaultendpunct}{\mcitedefaultseppunct}\relax
\EndOfBibitem
\bibitem[Baer \latin{et~al.}(2010)Baer, Livshits, and Salzner]{baer:10arpc}
Baer,~R.; Livshits,~E.; Salzner,~U. Tuned range-separated hybrids in density
  functional theory. \emph{Ann. Rev. Phys. Chem.} \textbf{2010}, \emph{61},
  85--109\relax
\mciteBstWouldAddEndPuncttrue
\mciteSetBstMidEndSepPunct{\mcitedefaultmidpunct}
{\mcitedefaultendpunct}{\mcitedefaultseppunct}\relax
\EndOfBibitem
\bibitem[Garrett \latin{et~al.}(2014)Garrett, Sosa~Vazquez, Egri, Wilmer,
  Johnson, Robinson, and Isborn]{garrett:14jctc}
Garrett,~K.; Sosa~Vazquez,~X.; Egri,~S.~B.; Wilmer,~J.; Johnson,~L.~E.;
  Robinson,~B.~H.; Isborn,~C.~M. Optimum exchange for calculation of excitation
  energies and hyperpolarizabilities of organic electro-optic chromophores.
  \emph{J. Chem. Theory Comput.} \textbf{2014}, \emph{10}, 3821--3831\relax
\mciteBstWouldAddEndPuncttrue
\mciteSetBstMidEndSepPunct{\mcitedefaultmidpunct}
{\mcitedefaultendpunct}{\mcitedefaultseppunct}\relax
\EndOfBibitem
\bibitem[Vuckovic \latin{et~al.}(2017)Vuckovic, Irons, Wagner, Teale, and
  Gori-Giorgi]{vuckovic:17pccp}
Vuckovic,~S.; Irons,~T. J.~P.; Wagner,~L.~O.; Teale,~A.~M.; Gori-Giorgi,~P.
  Interpolated energy densities, correlation indicators and lower bounds from
  approximations to the strong coupling limit of DFT. \emph{Phys. Chem. Chem.
  Phys.} \textbf{2017}, \emph{19}, 6169--6183\relax
\mciteBstWouldAddEndPuncttrue
\mciteSetBstMidEndSepPunct{\mcitedefaultmidpunct}
{\mcitedefaultendpunct}{\mcitedefaultseppunct}\relax
\EndOfBibitem
\bibitem[Via-Nadal \latin{et~al.}(2017)Via-Nadal, Rodr{\'i}guez-Mayorga, and
  Matito]{via-nadal:17pra}
Via-Nadal,~M.; Rodr{\'i}guez-Mayorga,~M.; Matito,~E. {A Salient Signature of
  van der Waals Interactions}. \emph{Phys. Rev. A} \textbf{2017}, \emph{96},
  050501\relax
\mciteBstWouldAddEndPuncttrue
\mciteSetBstMidEndSepPunct{\mcitedefaultmidpunct}
{\mcitedefaultendpunct}{\mcitedefaultseppunct}\relax
\EndOfBibitem
\bibitem[Coulson and Neilson(1961)Coulson, and Neilson]{coulson:61ppsl}
Coulson,~C.~A.; Neilson,~A.~H. Electron correlation in the ground state of
  helium. \emph{Proc. Phys. Soc. London} \textbf{1961}, \emph{78},
  831--837\relax
\mciteBstWouldAddEndPuncttrue
\mciteSetBstMidEndSepPunct{\mcitedefaultmidpunct}
{\mcitedefaultendpunct}{\mcitedefaultseppunct}\relax
\EndOfBibitem
\bibitem[Rodr\'iguez-Mayorga \latin{et~al.}(2017)Rodr\'iguez-Mayorga,
  Ramos-Cordoba, Via-Nadal, Piris, and Matito]{rodriguez:17pccp2}
Rodr\'iguez-Mayorga,~M.; Ramos-Cordoba,~E.; Via-Nadal,~M.; Piris,~M.;
  Matito,~E. Comprehensive benchmarking of density matrix functional
  approximations. \emph{Phys. Chem. Chem. Phys.} \textbf{2017}, \emph{19},
  24029--24041\relax
\mciteBstWouldAddEndPuncttrue
\mciteSetBstMidEndSepPunct{\mcitedefaultmidpunct}
{\mcitedefaultendpunct}{\mcitedefaultseppunct}\relax
\EndOfBibitem
\bibitem[Note41()]{Note42}
Rigorously speaking, the condition that the coefficients should satisfy is that
  their N$^{th}$ root, $N$ being the number of electrons, is close to one for
  the system to be dominated by dynamic correlation. Otherwise, we would not
  correctly consider as such composites of non-interacting
  dynamic-correlation-driven systems, \eg, an infinite number of
  non-interacting helium atoms\relax
\mciteBstWouldAddEndPuncttrue
\mciteSetBstMidEndSepPunct{\mcitedefaultmidpunct}
{\mcitedefaultendpunct}{\mcitedefaultseppunct}\relax
\EndOfBibitem
\bibitem[Kutzelnigg and Mukherjee(1999)Kutzelnigg, and
  Mukherjee]{kutzelnigg:99jcp}
Kutzelnigg,~W.; Mukherjee,~D. Cumulant expansion of the reduced density
  matrices. \emph{J. Chem. Phys.} \textbf{1999}, \emph{110}, 2800--2809\relax
\mciteBstWouldAddEndPuncttrue
\mciteSetBstMidEndSepPunct{\mcitedefaultmidpunct}
{\mcitedefaultendpunct}{\mcitedefaultseppunct}\relax
\EndOfBibitem
\bibitem[Ernzerhof \latin{et~al.}(1996)Ernzerhof, Burke, and
  Perdew]{ernzerhof:96jcp}
Ernzerhof,~M.; Burke,~K.; Perdew,~J.~P. Long-range asymptotic behavior of
  ground-state wave functions, one-matrices, and pair densities. \emph{J. Chem.
  Phys.} \textbf{1996}, \emph{105}, 2798--2803\relax
\mciteBstWouldAddEndPuncttrue
\mciteSetBstMidEndSepPunct{\mcitedefaultmidpunct}
{\mcitedefaultendpunct}{\mcitedefaultseppunct}\relax
\EndOfBibitem
\bibitem[Note41()]{Note41}
We have performed full configuration interaction (FCI) calculations with the
  aug-cc-pVDZ basis set for He$_2$ and H$_n$ ($n=2-8$), and single and double
  configuration interactions calculations with the aug-cc-pVTZ basis set for
  HeNe and HeAr molecules. For the isoelectronic series of He we have used an
  even-tempered basis set of 5s, 5p and 5d functions, optimized following the
  procedure described elsewhere~\cite{matito:10pccp}\relax
\mciteBstWouldAddEndPuncttrue
\mciteSetBstMidEndSepPunct{\mcitedefaultmidpunct}
{\mcitedefaultendpunct}{\mcitedefaultseppunct}\relax
\EndOfBibitem
\bibitem[Mercero \latin{et~al.}(2003)Mercero, Valderrama, and
  Ugalde]{mercero:03bch}
Mercero,~J.~M.; Valderrama,~E.; Ugalde,~J.~M. In \emph{Metal-Ligand
  Interactions}; Russo,~N., Salahub,~D.~R., Witko,~M., Eds.; Kluwer Academic
  Publishers: The Netherlands, 2003; pp 205--239\relax
\mciteBstWouldAddEndPuncttrue
\mciteSetBstMidEndSepPunct{\mcitedefaultmidpunct}
{\mcitedefaultendpunct}{\mcitedefaultseppunct}\relax
\EndOfBibitem
\bibitem[Carrascal \latin{et~al.}(2015)Carrascal, Ferrer, Smith, and
  Burke]{carrascal:15jpcm}
Carrascal,~D.; Ferrer,~J.; Smith,~J.~C.; Burke,~K. The Hubbard dimer: a density
  functional case study of a many-body problem. \emph{J. Phys.: Condens.
  Matter} \textbf{2015}, \emph{27}, 393001\relax
\mciteBstWouldAddEndPuncttrue
\mciteSetBstMidEndSepPunct{\mcitedefaultmidpunct}
{\mcitedefaultendpunct}{\mcitedefaultseppunct}\relax
\EndOfBibitem
\bibitem[Deur \latin{et~al.}(2017)Deur, Mazouin, and Fromager]{deur:17prb}
Deur,~K.; Mazouin,~L.; Fromager,~E. Exact ensemble density functional theory
  for excited states in a model system: Investigating the weight dependence of
  the correlation energy. \emph{Phys. Rev. B} \textbf{2017}, \emph{95},
  035120\relax
\mciteBstWouldAddEndPuncttrue
\mciteSetBstMidEndSepPunct{\mcitedefaultmidpunct}
{\mcitedefaultendpunct}{\mcitedefaultseppunct}\relax
\EndOfBibitem
\bibitem[Chakravorty \latin{et~al.}(1993)Chakravorty, Gwaltney, Davidson,
  Parpia, and p~Fischer]{chakravorty:93pra}
Chakravorty,~S.~J.; Gwaltney,~S.~R.; Davidson,~E.~R.; Parpia,~F.~A.;
  p~Fischer,~C.~F. Ground-state correlation energies for atomic ions with 3 to
  18 electrons. \emph{Phys. Rev. A} \textbf{1993}, \emph{47}, 3649\relax
\mciteBstWouldAddEndPuncttrue
\mciteSetBstMidEndSepPunct{\mcitedefaultmidpunct}
{\mcitedefaultendpunct}{\mcitedefaultseppunct}\relax
\EndOfBibitem
\bibitem[Ramos-Cordoba \latin{et~al.}(2015)Ramos-Cordoba, Lopez, Piris, and
  Matito]{ramos-cordoba:15jcp}
Ramos-Cordoba,~E.; Lopez,~X.; Piris,~M.; Matito,~E. H$_4$: A challenging system
  for natural orbital functional approximations. \emph{J. Chem. Phys.}
  \textbf{2015}, \emph{143}, 164112\relax
\mciteBstWouldAddEndPuncttrue
\mciteSetBstMidEndSepPunct{\mcitedefaultmidpunct}
{\mcitedefaultendpunct}{\mcitedefaultseppunct}\relax
\EndOfBibitem
\bibitem[Piris \latin{et~al.}(2008)Piris, Lopez, and Ugalde]{piris:08jcp2}
Piris,~M.; Lopez,~X.; Ugalde,~J. Correlation holes for the helium dimer.
  \emph{J. Chem. Phys.} \textbf{2008}, \emph{128}, 134102\relax
\mciteBstWouldAddEndPuncttrue
\mciteSetBstMidEndSepPunct{\mcitedefaultmidpunct}
{\mcitedefaultendpunct}{\mcitedefaultseppunct}\relax
\EndOfBibitem
\bibitem[Pauling and Wilson(1935)Pauling, and Wilson]{pauling:35book}
Pauling,~L.; Wilson,~E.~B. \emph{Introduction to quantum mechanics}; Dover
  Publications, Inc.: New York, 1935\relax
\mciteBstWouldAddEndPuncttrue
\mciteSetBstMidEndSepPunct{\mcitedefaultmidpunct}
{\mcitedefaultendpunct}{\mcitedefaultseppunct}\relax
\EndOfBibitem
\bibitem[Hermann \latin{et~al.}(2017)Hermann, DiStasio~Jr, and
  Tkatchenko]{hermann:17cr}
Hermann,~J.; DiStasio~Jr,~R.~A.; Tkatchenko,~A. First-Principles Models for van
  der Waals Interactions in Molecules and Materials: Concepts, Theory, and
  Applications. \emph{Chem. Rev.} \textbf{2017}, \emph{117}, 4714--4758\relax
\mciteBstWouldAddEndPuncttrue
\mciteSetBstMidEndSepPunct{\mcitedefaultmidpunct}
{\mcitedefaultendpunct}{\mcitedefaultseppunct}\relax
\EndOfBibitem
\bibitem[Piris and Ugalde(2014)Piris, and Ugalde]{piris:14ijqc}
Piris,~M.; Ugalde,~J.~M. Perspective on natural orbital functional theory.
  \emph{Int. J. Quantum Chem.} \textbf{2014}, \emph{114}, 1169--1175\relax
\mciteBstWouldAddEndPuncttrue
\mciteSetBstMidEndSepPunct{\mcitedefaultmidpunct}
{\mcitedefaultendpunct}{\mcitedefaultseppunct}\relax
\EndOfBibitem
\bibitem[Matito \latin{et~al.}(2010)Matito, Cioslowski, and
  Vyboishchikov]{matito:10pccp}
Matito,~E.; Cioslowski,~J.; Vyboishchikov,~S.~F. Properties of harmonium atoms
  from FCI calculations: Calibration and benchmarks for the ground state of the
  two-electron species. \emph{Phys. Chem. Chem. Phys.} \textbf{2010},
  \emph{12}, 6712\relax
\mciteBstWouldAddEndPuncttrue
\mciteSetBstMidEndSepPunct{\mcitedefaultmidpunct}
{\mcitedefaultendpunct}{\mcitedefaultseppunct}\relax
\EndOfBibitem
\end{mcitethebibliography}
\end{document}